\documentclass[aps,prb,twocolumn,groupedaddress,10pt,a4paper,showpacs]{revtex4}

\usepackage[dvips]{graphicx}
\usepackage[intlimits]{amsmath}
\usepackage{amssymb}
\usepackage{amsthm}
\usepackage{booktabs}
\usepackage{multirow}
\usepackage{slashbox}
\usepackage{mathrsfs}

\newcommand{\ud}{\,\mathrm{d}}
\DeclareMathAlphabet{\mathcal}{OMS}{cmsy}{m}{n}

\begin{document}

\author{Tomasz J. Antosiewicz}
\email{tomasza@chalmers.se}
\author{S. Peter Apell}
\author{Michael Z\"ach}
\author{Igor Zori\'c}
\author{Christoph Langhammer}
\email{clangham@chalmers.se}
\affiliation{Chalmers University of Technology, Department of Applied Physics, SE-412 96 G\"oteborg, Sweden}

\title{Oscillatory optical response of amorphous plasmonic nanoparticle arrays}

\date{\today}

\begin{abstract}
The optical response of metallic nanoparticle arrays is dominated by localized surface plasmon excitations and is the sum of individual particle contributions modified by inter-particle coupling depending on specific array geometry. Here we scrutinize how experimentally measured properties of large scale (30 mm$^{2}$) amorphous Au nanodisk arrays stem from single particle properties and their interaction. They give rise to a distinct oscillatory behavior of the plasmon peak position, full-width at half-maximum, and extinction efficiency which depends on the minimum particle center-to-center (CC) distance.
\end{abstract}


\maketitle

Strong coupling of light to metal nanoparticles via localized surface plasmons is one reason for the wide exploitation of nanosized metallic entities.\cite{Nmat_8_867_zayats, Sci_326_1091_langhammer, Nmat_9_205_atwater, NMat_10_911_linic} For many targeted uses of nanoplasmonic systems a key question is whether to operate with individual metallic structures \cite{NL_3_1057_vanDuyne, Nmat_2011_giessen} or to use ensembles in the form of periodic \cite{PRL_101_143902_barnes} or random arrays on a support.\cite{ASCNano_5_2535_langhammer} The optical properties of nanoplasmonic arrays, both periodic and fully random, stem from the optical response of individual particles. The array modifies these single-particle spectra, sometimes quite considerably, via inter-particle coupling that depends on the exact array geometry. Thus, array design generally is an additional handle for tuning plasmonic response, together with particle size, geometry, and materials.

Here we scrutinize experimentally and theoretically a novel oscillatory behavior of the optical response from a specific type of nanoparticle array, somewhere between perfectly periodic and fully random, that we refer to as an \emph{amorphous} array. This particular type of nanoparticle arrangement on a surface exhibits short range distance order, while, at long distances, it is completely random. Furthermore, it can be quite easily fabricated on large areas (wafer scale), using bottom-up self-assembly based nanofabrication techniques like hole-mask colloidal lithography,\cite{AdvMat_19_4297_langhammer} making it a first choice for many large-scale devices and applications.\cite{Sci_326_1091_langhammer, ACSNano_5_6218_hagglund, NL_8_3155_vogelgesang, PRL_104_147401_sepulveda} Our finding shows, in contrast to the generally accepted opinion, that amorphous arrays exhibit distinct properties of interacting particles even if their density is low.

\begin{figure}
\centering
\includegraphics[width=8.5cm]{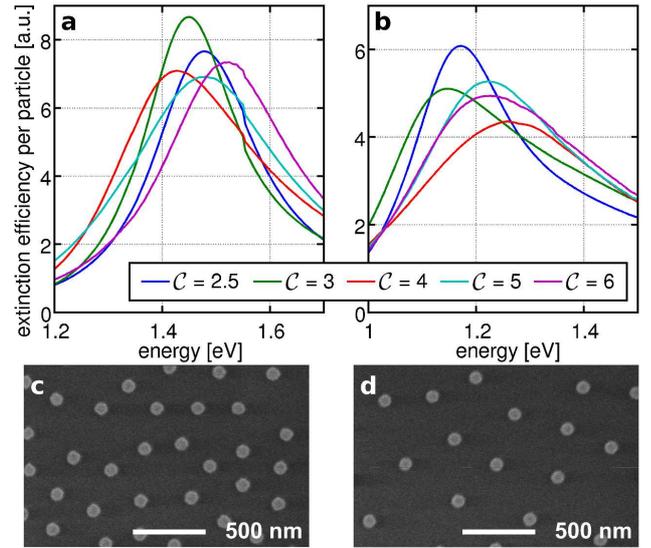}
\caption{(color online) Experimentally measured extinction spectra of gold nanodisks with an engineered randomness. Resonance position, peak value, and linewidth all show a non-monotonic dependence on lattice parameter $\mathcal{C}$ being minimum center-to-center distance in units of disk diameter $D$: (a) $D=160$ nm, (b) $260$ nm. (c) and (d) show SEM images of the amorphous arrays for $\mathcal{C}=3$ and 4, respectively. Notice the random distribution of perfectly defined particles.}
\label{fig::measurements}
\end{figure}

For our experiments, we fabricated large area arrays of gold nanodisks with engineered randomness using an electron beam-lithography (EBL) nanofabrication scheme.\cite{ASCNano_5_2535_langhammer} Circular areas of roughly 30 mm$^{2}$ were patterned for each considered center-to-center (CC) distance with minimum imposed CC ranging from 2.5 to 7 in units of particle diameter $D$ ($\mathcal{C}$). The size distribution for the disks is very narrow as can be seen in Fig. \ref{fig::measurements} and the histograms in Ref. \cite{ASCNano_5_2535_langhammer} and, consequently, basically eliminates inhomogeneous broadening. We fabricated large arrays (30 mm$^{2}$) with a very large number of particles to eliminate the effect that for the same set of global parameters (minimum CC, $D$, thickness 20 nm, illumination conditions fixed) small samples could correspond to slightly different array realizations.

In the first two panels of Fig. \ref{fig::measurements} we present experimentally measured extinction efficiency spectra near the resonance to illustrate the sensitivity of the peak position and extinction efficiency per particle at peak to the CC value $\mathcal{C}$ for $D=160$ nm in (a) and $D=260$ nm in (b). In a first rough analysis for $D=160$ nm we see, that for $\mathcal{C}=2.5$ extinction is maximal at 838 nm, then it redshifts for $\mathcal{C}=3$ and 4 to 855 and 867 nm, respectively, and then undergoes a blueshift to 837 nm ($\mathcal{C}=5$) and 815 nm for $\mathcal{C}=6$ -- suggesting an oscillatory behavior of the peak position. A similar trend is also seen when tracing the peak amplitude (extinction efficiency) and the linewidth.

An efficient way to model arrays of plasmonic nanoparticles is by a coupled dipole approximation in which each disk is modeled by an induced point dipole coupled to an external electromagnetic field.\cite{JOSAA_11_1491_draine} In this framework, the particle properties are described by a polarizability $\alpha$ determined by the material, geometry, and surrounding medium.\cite{bohren_huffman, JOSAB_26_517_moroz} In the quasistatic regime the polarizability $\alpha_{\mathrm{qs}}$ is proportional to $V(\epsilon_{m}-\epsilon_{s})/(\epsilon_{s}+L(\epsilon_{m}-\epsilon_{s}))$, where $\epsilon_{m}$ and $\epsilon_{s}$ are the permittivities of the metal particle and surrounding medium, respectively, $V$ is the particle volume, and $L$ is a shape depolarization factor. Dynamic depolarization and radiative damping are accounted for by introducing the modified long wavelength approximation \cite{JOSAB_26_517_moroz,JCS_10_295_schatz} $1/\alpha=1/\alpha_{\mathrm{qs}}-\frac{2}{3}ik^{3}-\frac{k^2}{a}$, where $k$ is the wave number of exciting light of wavelength $\lambda$ and $a$ is a length associated with the particle geometry.\cite{JOSAB_26_517_moroz}

For an infinite periodic array, where the particles are interacting, the system of coupled equations is solved by assuming that the polarization of each particle is the same and thus $\alpha$ becomes an effective polarizability that takes into account inter-particle interactions via a retarded dipole sum.\cite{PRL_101_143902_barnes,JCP_121_12606_schatz,JPhysB_38_L115_markel} However, with a gradual increase of disorder, the narrow peak characteristic for a periodic array disappears and the response turns into an inhomogeneously broadened plasmon resonance.\cite{OL_34_401_barnes}

For an \emph{amorphous} array we can, as an ensemble average, define an effective polarization $\alpha^{*}$. One way of analyzing the inter-particle contributions to $\alpha^{*}$ is to average over many realizations of amorphous arrays (\emph{i.e.} dipole sums). Here, however, we describe a model in which the average particle is surrounded by a continuous \emph{film} of dipoles with surface densities determined by the pair correlation function $\mathcal{G}(r,\mathcal{C})$, where $r$ is the radial distance from the considered particle. One can think of this as an average of an infinite number of different realizations of amorphous arrays centered around a specified particle placed at a particular specified point. In a sense this approach is reminiscent of the coherent-potential approximation for a random distribution of particles on a square array.\cite{PRB_28_4247_persson}

For the average particle in an amorphous array we carry out the same procedure of solving the discrete dipole equations as in \cite{JCP_121_12606_schatz} where the retarded dipole sum (discrete particles) is replaced by a retarded dipole integral (continuous film with hole)
\begin{multline}\mathcal{S}(\mathcal{C})= \int_{\ell_{\mathrm{cc}}}^{+\infty}\int_{0}^{2\pi}e^{ikr}\left[ \frac{\left(1-ikr\right)\left(3\cos^{2}\theta-1\right)}{r^{3}} + \right.\\ \left. + \frac{k^{2}\sin^{2}\theta}{r} \right]g(r,\mathcal{C})\ r\ud \theta\ud r, \label{eq::S} \end{multline}
where the exponential term multiplied by the expression in the square brackets ($e^{ikr}[\dots]$) describes the retarded dipole-dipole interaction, $g(r,\mathcal{C})=\sigma \mathcal{G}(r,\mathcal{C})$ is the pair correlation function $\mathcal{G}(r,\mathcal{C})$ multiplied by the particle surface density $\sigma=\sigma_{0}\ell_{\mathrm{cc}}^{-2}$, $\ell_{\mathrm{cc}}\equiv\mathcal{C}D$, and $\sigma_{0}$ is a surface packing parameter. The integration is over the whole 2D $(r,\theta)$ space with the exception of an inner circle smaller than $\ell_{\mathrm{cc}}$. Performing the angular average yields the average, effective polarizability \cite{JCP_121_12606_schatz,PRL_101_143902_barnes}
\begin{equation}
\alpha^{*}=\frac{1}{\alpha^{-1} - \mathcal{S}},
\label{eq::alpha_star}
\end{equation}
where
\begin{equation}
\mathcal{S}= \pi\sigma\int_{\ell_{\mathrm{cc}}}^{+\infty} e^{ikr}\left(k^{2} + \frac{1-ikr}{r^{2}}\right) \mathcal{G}(r,\mathcal{C})\ud r.
\label{eq::S_1}
\end{equation}
The function $e^{ikr}(k^2+(1-ikr)/r^{2})$ consists of two parts: the first ($e^{ikr}k^{2}$) comes from the far-field dipole radiation and its value oscillates, while the second ($e^{ikr}(1-ikr)/r^{2}$) corresponds to intermediate and near-fields and its value has a well-defined limit for $r\to\infty$. However, these observations are only strictly valid for a well behaved function $\mathcal{G}(r,\mathcal{C})$ at infinity.

To perform the integration for $\mathcal{S}$ in Eq. \ref{eq::S_1} we require an expression for $\mathcal{G}(\rho,\mathcal{C})$, where $\rho=r/D$ is a normalized radius. We obtain $\mathcal{G}(\rho,\mathcal{C})$ by finding a function which fits well ($R^{2}\simeq1$) to pair correlation data calculated from random distributions of particles generated with the random sequential adsorption algorithm, which was also used to calculate particle positions for the fabricated arrays.\cite{JStatPhys_44_793_feder} The selected function consists of two parts --  a constant and a varying one
\begin{multline}\label{eq::fit_G}\mathcal{G}(\rho,\mathcal{C})= 1 + \sin\left(2\pi\frac{\rho-d_{0}\mathcal{C}}{d_{1}\mathcal{C}}\right)\left[ a_{0} e^{-a_{1}\mathcal{C}^{a_{2}}(\rho-c\mathcal{C})} +\right. \\ \left. + b_{0}e^{-b_{1}\mathcal{C}} e^{-b_{2}\mathcal{C}^{b_{3}}(\rho-c\mathcal{C})} \right] \ \ \mathrm{for}\ \rho\ge\mathcal{C}.\end{multline}
It is chosen because its product with functions describing dipole fields is relatively easy to compute and its $R^{2}=0.99$ (fitting parameters given in \cite{ZZZ_cc_fitting}). Slight differences between this function and the one shown by Hinrichsen \emph{et al.} \cite{JStatPhys_44_793_feder} occur only for particles at close distances. However, as we show later, a qualitative description is insensitive to the exact expression describing the short range order.

Equation \ref{eq::fit_G}, while easily integratable when multiplied by the expression for dipole radiation does not lend itself to an easy exposition of the main physics taking place. Therefore a qualitative analysis is carried out first, before performing the full calculation. We do this by keeping the hard-core part (unity) and omitting the second, oscillating term (multiplied by the sine function). Thus, the simplified $\mathcal{G}$ reduces to a Heaviside step function $\Theta(\rho-\mathcal{C})$ that describes a fully random array with a removed circle of radius $\mathcal{C}$ around the average particle. The simplification still describes satisfyingly the most important property of the array, namely the short-range order defined by the minimal allowed CC distance for the analyzed particle, and does not alter the main physical processes occurring within the array.

The function $\mathcal{G}=\Theta(\rho-\mathcal{C})$ is derived from Eq. \ref{eq::fit_G} by setting the pair correlation function to unity. To calculate the far-field term, which oscillates around a mean value, we modify it by adding to the exponent the term $-\varepsilon r$, which makes the expression $\pi\sigma k^{2}\int_{\ell_{\mathrm{cc}}}^{+\infty} e^{ikr-\varepsilon r} \ud r$ well-defined. This represents a case when the array is illuminated by a very broad Gaussian beam, \emph{i.e.} disks very far away from the center of the beam feel a diminished intensity of the electric field, but the decay is slow enough that the average treatment of the disks holds.

The first integral in $\lim_{\varepsilon\to0}$ equals $\pi\sigma kie^{ik\ell_{\mathrm{cc}}}$ and the second $\pi\sigma e^{ik\ell_{\mathrm{cc}}}/\ell_{\mathrm{cc}}$. Thus, for the simplified case of $\mathcal{G}=\Theta(\rho-\mathcal{C})$, the retarded dipole integral becomes
\begin{equation}
\mathcal{S}^{\Theta}=\pi\sigma \frac{e^{ik\ell_{\mathrm{cc}}}}{\ell_{\mathrm{cc}}}\left(1+ik\ell_{\mathrm{cc}}\right).
\label{eq::Stheta}
\end{equation}
We substitute this result into the effective polarizability $\alpha^{*}$ (Eq. \ref{eq::alpha_star}) and rewrite the right hand side containing the substituted expressions into a Lorentzian form to easily identify the peak position and full-width at half-maximum (FWHM). We employ, as an illustrative test case, a generic metal sphere with a Drude dielectric function $\epsilon(\omega)=1-\omega^{2}_{p}/(\omega(\omega+i\gamma))$. Using the modified long wavelength approximation\cite{JOSAB_26_517_moroz,JCS_10_295_schatz} we get
\begin{multline}
\frac{\alpha^{*}}{4\pi\epsilon_{0}R^{3}}=\\
=\frac{1-\bar{\omega}^{2}(1+s^{2})-qf+i(\bar{\omega}(\bar{\gamma}+\frac{2}{3}s^{3}\bar{\omega}^{2})+qg)}{(1-\bar{\omega}^{2}(1+s^{2})-qf)^{2}+(\bar{\omega}(\bar{\gamma}+\frac{2}{3}s^{3}\bar{\omega}^{2})+qg)^{2}}, \label{eq::res_cond}
\end{multline}
where we have introduced dimensionless variables $\bar{\omega}\equiv\frac{\omega}{\omega_{0}}$, $\bar{\gamma}\equiv\frac{\gamma}{\omega_{0}}$, $s\equiv\frac{\omega_{0}R}{c}$ with $\omega_{0}^{2}=\frac{\omega_{p}^{2}}{3}$ the Mie resonance frequency, a coupling strength $q\equiv\pi\sigma_{0}(R/\ell_{\mathrm{cc}})^{3}$ ($q$ is maximum $\frac{\pi}{4}$ for $\sigma_{0}=1$), two functions $f(k\ell_{\mathrm{cc}})$ and $g(k\ell_{\mathrm{cc}})$: $f(x)=\cos{x}-x\sin{x}$, $g(x)=\sin{x}+x\cos{x}$, and $c$ is the speed of light.

\begin{figure}
\centering
\includegraphics[width=8.5cm]{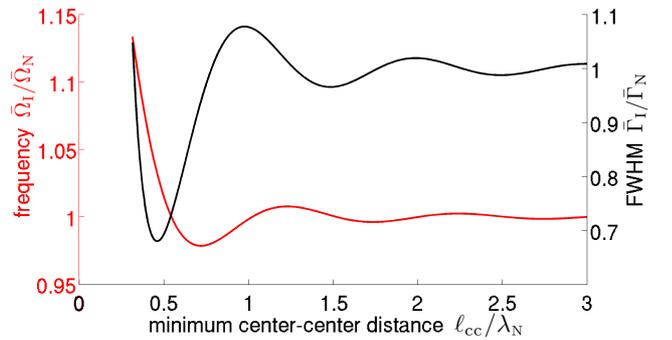}
\caption{(color online) The variation of the optical characteristics of the particle arrays in Fig.~\ref{fig::measurements} can be understood from a simple generic Drude model. The resonance frequency (red) and FWHM (black) of the array using a hard core pair-correlation function with minimum center-center distance $\ell_{\mathrm{cc}}$ is shown. We measure the center-center distance $\ell_{\mathrm{cc}}$ in units of the resonance wavelength $\lambda_{\mathrm{N}}$. The resonance frequency as well as the linewidth are in units of the bare particle properties. Notice how the interference between the particles causes both resonance frequency and FWHM to oscillate and the oscillations of the FWHM are larger than for the resonance frequency. With decreased coupling the curves settle to the particle values.}
\label{fig::alfa}
\end{figure}

From the denominator of Eq. (\ref{eq::res_cond}) we can read off the resonance frequency and the FWHM. In the non-interacting case ($q=0$) we have a mode at $\bar{\Omega}_{\mathrm{N}}=\frac{1}{\sqrt{1+s^{2}}}$ with a linewidth $\bar{\Gamma}_{\mathrm{N}}=\frac{2}{\sqrt{1+s^{2}}}(\bar{\gamma}+\frac{2s^{3}}{3(1+s^{2})})$. For an amorphous array consisting of such particles the resonance is modified by inter-particle coupling via the retarded dipole integral and Eq. (\ref{eq::res_cond}) shows that we have a resonance at
\begin{equation}
\bar{\Omega}_{\mathrm{I}}/\bar{\Omega}_{\mathrm{N}}=\sqrt{1-q\,f(k_{\mathrm{N}}\ell_{\mathrm{cc}})} \label{eq::omgea}
\end{equation}
in terms of the non-interacting $\bar{\Omega}_{\mathrm{N}}$ and a linewidth of
\begin{equation}
\bar{\Gamma}_{\mathrm{I}}/\bar{\Gamma}_{\mathrm{N}}=1+2(1+s^{2}) Q_{\mathrm{N}}g(k_{\mathrm{N}}\ell_{\mathrm{cc}}),   \label{eq::gamma}
\end{equation}
where $Q_{\mathrm{N}}$ is the individual particle quality factor. We see from Eqs. (\ref{eq::omgea}) and (\ref{eq::gamma}) that due to the quality factor we expect the randomness to show up much more in the FWHM than in the resonance frequency. This is clearly seen in Fig. \ref{fig::alfa} which shows peak position (red) and FWHM (black) based on Eqs. (\ref{eq::omgea}) and (\ref{eq::gamma}). The peak position follows a sinusoidal line with a period determined by the ratio of the minimum particle-particle distance to the resonance wavelength $\lambda_{\mathrm{N}}$ of a single particle. Notice, that the oscillations are governed by the low cut-off $\ell_{\mathrm{cc}}$. The real part of the dipole interaction term shifts the resonance position towards higher frequency when $\Re(\mathcal{S})\equiv f<0$, while for $\Re(\mathcal{S})>0$ it induces a red shift. The largest shifts occur when the array is relatively dense, \emph{cf.} large coupling strength $q$. For large inter-particle spacing the interference between the disks vanishes since $q$ is proportional to $\ell_{\mathrm{cc}}^{-3}$. 

The single particle resonance linewidth $\bar{\Gamma}_{\mathrm{I}}$ is modified by the imaginary part of $\mathcal{S}^{\Theta}\equiv g$ which introduces a modulation of the linewidth equal to $(1+s^{2}) Q_{\mathrm{N}}g(k_{\mathrm{N}}\ell_{\mathrm{cc}})$. Similar to the peak position, the linewidth is a decaying oscillatory function that tends to the single particle resonance width for infinitely diluted amorphous arrays. When $g<0$ for even-numbered half-periods the resonance linewidth is smaller than the single particle one, however, it does not go to zero.\cite{JOSAB_26_517_moroz, JCP_120_10871_schatz, JCP_122_097101_markel, JCP_122_097102_schatz} Note, that in this analysis we have kept the minimum center-center distance larger than of the order of two diameters so that higher order multipoles which are not present in the theory should be negligible.

In the above \emph{qualitative} picture, in which we dropped the oscillatory term in the pair correlation function, we have shown that the oscillations of the optical cross sections of amorphous arrays are the result of interference between the incident field driving a particle and the scattered fields originating from the other particles in the array. We use now the full expression for $\mathcal{G}$ (Eq. \ref{eq::fit_G}) to analyze experimental data, shown in Fig. \ref{fig::measurements}, obtained from extinction measurements on nanofabricated amorphous arrays of nanodisks. To model the properties of a single disk in the array, we adjust an oblate spheroidal polarizability so that the single disk (for a hypothetical case of an infinitely diluted array) resonance position and linewidth correspond to the asymptotes of the experimental arrays for very large CC.

\begin{figure}
\centering
\includegraphics[width=8.5cm]{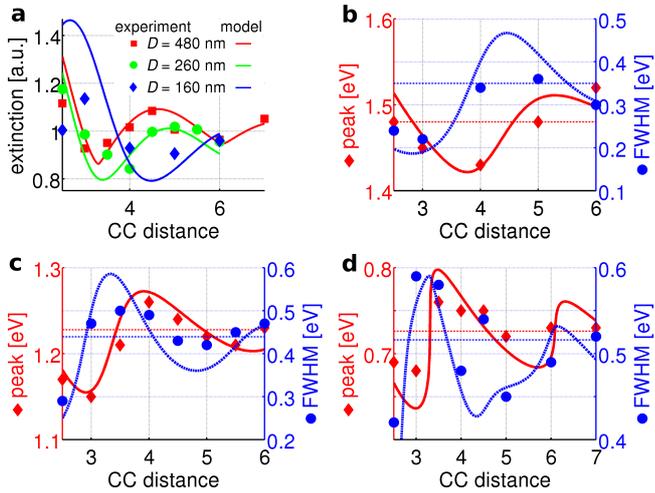}
\caption{(color online) Experimental results compared to our predictions based on the full pair-correlation function for the extinction at peak value, the peak position and linewidth as a function of the minimum particle-particle distance in units of particle diameter. The characteristics of the spectra (extinction value, position, and linewidth) oscillate as a result of radiative coupling between the plasmonic particles within the array and are a function of the minimum CC distance. (a) shows peak extinction values for a disk in the amorphous array for the different disk diameters, which are normalized to the peak extinction values of their respective single particles; (b)-(d) show extinction peak position (red left y-axis) and linewidth (blue right y-axis) for diameters (b) $160$, (c) $260$, and (d) $480$ nm as a function of the CC distance. The horizontal thin dash-dotted lines indicate asymptotes of the peak position and linewidth of the amorphous arrays.}
\label{fig::resultsDALL}
\end{figure}

First, we address the extinction cross sections per particle in the arrays. Figure \ref{fig::resultsDALL}a presents the measured and calculated extinction per disk in the array, $C_{\mathrm{e}}$, normalized to the extinction of a single disk. Clearly, extinction exhibits strong oscillatory CC distance dependence for all three measured particle sizes. The minimum-to-maximum difference is about 40\% for small CC distances and agrees very well with the model calculations.

The measured experimental peak positions (diamonds) and full-width at half-maximum (FWHM, circles) values are shown in Fig. \ref{fig::resultsDALL}b-d. The solid red and dashed blue lines representing peak position and FWHM values, respectively, are calculated according to the scheme outlined above. The agreement between the theoretical and experimental data is very good, in particular in view of our relatively simple theoretical treatment of the particle interaction in the array. The period and phase shift of the measured and calculated oscillations of the peak position and FWHM are consistent and show a pronounced CC distance dependence. Notably, as the most extreme case, the experimentally measured FWHM for the $D=260$ nm disk varies between \emph{ca.} 0.3 and 0.5 eV at a rather moderate change of the peak position (1.15 to 1.25 eV).

We now briefly address trends in the amplitude of the oscillations that physically originate from radiative coupling as a function of nanodisk size in the amorphous array. It is known from theory and experiment that the scattering efficiency, especially compared with absorption (\emph{i.e.} the scattering/absorption branching ratio), increases with particle size.\cite{JCP_126_194702_langhammer}  Thus, it is to be expected that the oscillations are most pronounced for the largest particles, where the radiative coupling is strong, and decrease in amplitude as the nanodisk diameter decreases. This is precisely the trend that can be seen in Fig. \ref{fig::resultsDALL} (for $D=480$ nm the maximum peak position oscillation amplitude ($\Delta E$) normalized to peak position ($E_{0}$) is $\Delta E/E_{0}=0.11$, for $D=260$ nm $\Delta E/E_{0}=0.09$, and for $D=160$ nm $\Delta E/E_{0}=0.06$). Consequently, the decrease of the oscillatory amplitude is expected to continue as the nanoparticle diameter further decreases. This can be clearly seen in Fig. \ref{fig::alfa}, where a 60 nm Drude sphere was considered.

The observed discrepancies between the calculated values for the peak position and, in particular, the FWHM are the result of several factors, among which are inhomogeneity of the fabricated disks (this effect is, however, small), the fact that disks with large diameter-to-thickness ratios are not perfectly described by one dipole, and the estimation of the long-range interference term in the dipole integral. Furthermore, higher order terms also start to become important. Another issue to be noted here is the illumination of the array in the modeling, \emph{i.e.} the assumption of a very broad Gaussian beam incident onto an infinite amorphous array (the array is in focus, so the phase of the incident beam is uniform). The latter assumption is then used to calculate the far-field term. However, for a finite array this may not be fully correct. Far-field radiation is proportional to $e^{ikr}$ and its definite integral (present in the retarded dipole integral) oscillates, so the value of the sum for a finite array depends on the exact relation between the array and the illumination. Thus, our result should be viewed as an average value which lies between a minimum and maximum and may be a little off for some arrays.

In summary, we have shown experimentally and explained, using a dipolar model, an oscillatory optical response of amorphous plasmonic nanoparticle arrays that depends on the minimum allowed particle-particle separation. Optical spectra of amorphous arrays, while stemming from those of single particles, exhibit a strong influence of intra-array radiative coupling of the plasmonic disks that results in oscillation of the extinction, its position and linewidth.

We acknowledge support from the Swedish Foundation for Strategic Research via project SSF RMA08, the Foundation for Strategic Environmental Research (Mistra Dnr 2004-118), the Swedish Energy Agency project 32078-1, the Formas project 229-2009-772, and the Swedish Research Council project 2010-4041.


\begin{thebibliography}{26}
\expandafter\ifx\csname natexlab\endcsname\relax\def\natexlab#1{#1}\fi
\expandafter\ifx\csname bibnamefont\endcsname\relax
  \def\bibnamefont#1{#1}\fi
\expandafter\ifx\csname bibfnamefont\endcsname\relax
  \def\bibfnamefont#1{#1}\fi
\expandafter\ifx\csname citenamefont\endcsname\relax
  \def\citenamefont#1{#1}\fi
\expandafter\ifx\csname url\endcsname\relax
  \def\url#1{\texttt{#1}}\fi
\expandafter\ifx\csname urlprefix\endcsname\relax\def\urlprefix{URL }\fi
\providecommand{\bibinfo}[2]{#2}
\providecommand{\eprint}[2][]{\url{#2}}

\bibitem[{\citenamefont{Kabashin et~al.}(2009)\citenamefont{Kabashin, Evans,
  Pastkovsky, Hendren, Wurtz, Atkinson, Pollard, Podolskiy, and
  Zayats}}]{Nmat_8_867_zayats}
\bibinfo{author}{\bibfnamefont{A.~V.} \bibnamefont{Kabashin}},
  \bibinfo{author}{\bibfnamefont{P.}~\bibnamefont{Evans}},
  \bibinfo{author}{\bibfnamefont{S.}~\bibnamefont{Pastkovsky}},
  \bibinfo{author}{\bibfnamefont{W.}~\bibnamefont{Hendren}},
  \bibinfo{author}{\bibfnamefont{G.~A.} \bibnamefont{Wurtz}},
  \bibinfo{author}{\bibfnamefont{R.}~\bibnamefont{Atkinson}},
  \bibinfo{author}{\bibfnamefont{R.}~\bibnamefont{Pollard}},
  \bibinfo{author}{\bibfnamefont{V.~A.} \bibnamefont{Podolskiy}},
  \bibnamefont{and} \bibinfo{author}{\bibfnamefont{A.~V.}
  \bibnamefont{Zayats}}, \bibinfo{journal}{Nature Mater.}
  \textbf{\bibinfo{volume}{8}}, \bibinfo{pages}{867} (\bibinfo{year}{2009}).

\bibitem[{\citenamefont{Larsson et~al.}(2009)\citenamefont{Larsson, Langhammer,
  Zori\'c, and Kasemo}}]{Sci_326_1091_langhammer}
\bibinfo{author}{\bibfnamefont{E.~M.} \bibnamefont{Larsson}},
  \bibinfo{author}{\bibfnamefont{C.}~\bibnamefont{Langhammer}},
  \bibinfo{author}{\bibfnamefont{I.}~\bibnamefont{Zori\'c}}, \bibnamefont{and}
  \bibinfo{author}{\bibfnamefont{B.}~\bibnamefont{Kasemo}},
  \bibinfo{journal}{Science} \textbf{\bibinfo{volume}{326}},
  \bibinfo{pages}{1091} (\bibinfo{year}{2009}).

\bibitem[{\citenamefont{Atwater and Polman}(2010)}]{Nmat_9_205_atwater}
\bibinfo{author}{\bibfnamefont{H.~A.} \bibnamefont{Atwater}} \bibnamefont{and}
  \bibinfo{author}{\bibfnamefont{A.}~\bibnamefont{Polman}},
  \bibinfo{journal}{Nature Mater.} \textbf{\bibinfo{volume}{9}},
  \bibinfo{pages}{205} (\bibinfo{year}{2010}).

\bibitem[{\citenamefont{Linic et~al.}(2011)\citenamefont{Linic, Christopher,
  and Ingham}}]{NMat_10_911_linic}
\bibinfo{author}{\bibfnamefont{S.}~\bibnamefont{Linic}},
  \bibinfo{author}{\bibfnamefont{P.}~\bibnamefont{Christopher}},
  \bibnamefont{and} \bibinfo{author}{\bibfnamefont{D.~B.}
  \bibnamefont{Ingham}}, \bibinfo{journal}{Nature Mater.}
  \textbf{\bibinfo{volume}{10}}, \bibinfo{pages}{911} (\bibinfo{year}{2011}).

\bibitem[{\citenamefont{Mc{F}arland and Van~Duyne}(2003)}]{NL_3_1057_vanDuyne}
\bibinfo{author}{\bibfnamefont{A.~D.} \bibnamefont{Mc{F}arland}}
  \bibnamefont{and} \bibinfo{author}{\bibfnamefont{R.~P.}
  \bibnamefont{Van~Duyne}}, \bibinfo{journal}{Nano Lett.}
  \textbf{\bibinfo{volume}{3}}, \bibinfo{pages}{1057} (\bibinfo{year}{2003}).

\bibitem[{\citenamefont{Liu et~al.}(2011)\citenamefont{Liu, Tang, Hentschel,
  Giessen, and Alivisatos}}]{Nmat_2011_giessen}
\bibinfo{author}{\bibfnamefont{N.}~\bibnamefont{Liu}},
  \bibinfo{author}{\bibfnamefont{M.~L.} \bibnamefont{Tang}},
  \bibinfo{author}{\bibfnamefont{M.}~\bibnamefont{Hentschel}},
  \bibinfo{author}{\bibfnamefont{H.}~\bibnamefont{Giessen}}, \bibnamefont{and}
  \bibinfo{author}{\bibfnamefont{A.~P.} \bibnamefont{Alivisatos}},
  \bibinfo{journal}{Nature Mater.} \textbf{\bibinfo{volume}{10}},
  \bibinfo{pages}{631} (\bibinfo{year}{2011}).

\bibitem[{\citenamefont{Augui\'e and Barnes}(2008)}]{PRL_101_143902_barnes}
\bibinfo{author}{\bibfnamefont{B.}~\bibnamefont{Augui\'e}} \bibnamefont{and}
  \bibinfo{author}{\bibfnamefont{W.~L.} \bibnamefont{Barnes}},
  \bibinfo{journal}{Phys. Rev. Lett.} \textbf{\bibinfo{volume}{101}},
  \bibinfo{pages}{143902} (\bibinfo{year}{2008}).

\bibitem[{\citenamefont{Zori\'c et~al.}(2011)\citenamefont{Zori\'c, Z\"ach,
  Kasemo, and Langhammer}}]{ASCNano_5_2535_langhammer}
\bibinfo{author}{\bibfnamefont{I.}~\bibnamefont{Zori\'c}},
  \bibinfo{author}{\bibfnamefont{M.}~\bibnamefont{Z\"ach}},
  \bibinfo{author}{\bibfnamefont{B.}~\bibnamefont{Kasemo}}, \bibnamefont{and}
  \bibinfo{author}{\bibfnamefont{C.}~\bibnamefont{Langhammer}},
  \bibinfo{journal}{ACS Nano} \textbf{\bibinfo{volume}{5}},
  \bibinfo{pages}{2535} (\bibinfo{year}{2011}).

\bibitem[{\citenamefont{Fredriksson et~al.}(2007)\citenamefont{Fredriksson,
  Alaverdyan, Dmitriev, Langhammer, Sutherland, Z\"ach, and
  Kasemo}}]{AdvMat_19_4297_langhammer}
\bibinfo{author}{\bibfnamefont{H.}~\bibnamefont{Fredriksson}},
  \bibinfo{author}{\bibfnamefont{Y.}~\bibnamefont{Alaverdyan}},
  \bibinfo{author}{\bibfnamefont{A.}~\bibnamefont{Dmitriev}},
  \bibinfo{author}{\bibfnamefont{C.}~\bibnamefont{Langhammer}},
  \bibinfo{author}{\bibfnamefont{D.~S.} \bibnamefont{Sutherland}},
  \bibinfo{author}{\bibfnamefont{M.}~\bibnamefont{Z\"ach}}, \bibnamefont{and}
  \bibinfo{author}{\bibfnamefont{B.}~\bibnamefont{Kasemo}},
  \bibinfo{journal}{Adv. Mater.} \textbf{\bibinfo{volume}{19}},
  \bibinfo{pages}{4297} (\bibinfo{year}{2007}).

\bibitem[{\citenamefont{Gusak et~al.}(2011)\citenamefont{Gusak, Kasemo, and
  H\"agglund}}]{ACSNano_5_6218_hagglund}
\bibinfo{author}{\bibfnamefont{V.}~\bibnamefont{Gusak}},
  \bibinfo{author}{\bibfnamefont{B.}~\bibnamefont{Kasemo}}, \bibnamefont{and}
  \bibinfo{author}{\bibfnamefont{C.}~\bibnamefont{H\"agglund}},
  \bibinfo{journal}{ACS Nano} \textbf{\bibinfo{volume}{5}},
  \bibinfo{pages}{6218} (\bibinfo{year}{2011}).

\bibitem[{\citenamefont{Esteban et~al.}(2008)\citenamefont{Esteban,
  Vogelgesang, Dorfm\"uller, Dmitriev, Rockstuhl, Etrich, and
  Kern}}]{NL_8_3155_vogelgesang}
\bibinfo{author}{\bibfnamefont{R.}~\bibnamefont{Esteban}},
  \bibinfo{author}{\bibfnamefont{R.}~\bibnamefont{Vogelgesang}},
  \bibinfo{author}{\bibfnamefont{J.}~\bibnamefont{Dorfm\"uller}},
  \bibinfo{author}{\bibfnamefont{A.}~\bibnamefont{Dmitriev}},
  \bibinfo{author}{\bibfnamefont{C.}~\bibnamefont{Rockstuhl}},
  \bibinfo{author}{\bibfnamefont{C.}~\bibnamefont{Etrich}}, \bibnamefont{and}
  \bibinfo{author}{\bibfnamefont{K.}~\bibnamefont{Kern}},
  \bibinfo{journal}{Nano Lett.} \textbf{\bibinfo{volume}{8}},
  \bibinfo{pages}{3155} (\bibinfo{year}{2008}).

\bibitem[{\citenamefont{Sep\'ulveda et~al.}(2010)\citenamefont{Sep\'ulveda,
  Gonz\'alez-D\'iaz, Garc\'ia-Mart\'in, Lechuga, and
  Armelles}}]{PRL_104_147401_sepulveda}
\bibinfo{author}{\bibfnamefont{B.}~\bibnamefont{Sep\'ulveda}},
  \bibinfo{author}{\bibfnamefont{J.~B.} \bibnamefont{Gonz\'alez-D\'iaz}},
  \bibinfo{author}{\bibfnamefont{A.}~\bibnamefont{Garc\'ia-Mart\'in}},
  \bibinfo{author}{\bibfnamefont{L.~M.} \bibnamefont{Lechuga}},
  \bibnamefont{and} \bibinfo{author}{\bibfnamefont{G.}~\bibnamefont{Armelles}},
  \bibinfo{journal}{Phys. Rev. Lett.} \textbf{\bibinfo{volume}{104}},
  \bibinfo{pages}{147401} (\bibinfo{year}{2010}).

\bibitem[{\citenamefont{Draine and Flatau}(1994)}]{JOSAA_11_1491_draine}
\bibinfo{author}{\bibfnamefont{B.~T.} \bibnamefont{Draine}} \bibnamefont{and}
  \bibinfo{author}{\bibfnamefont{P.~J.} \bibnamefont{Flatau}},
  \bibinfo{journal}{J. Opt. Soc. Am. A} \textbf{\bibinfo{volume}{11}},
  \bibinfo{pages}{1491} (\bibinfo{year}{1994}).

\bibitem[{\citenamefont{Bohren and Huffman}(1983)}]{bohren_huffman}
\bibinfo{author}{\bibfnamefont{C.}~\bibnamefont{Bohren}} \bibnamefont{and}
  \bibinfo{author}{\bibfnamefont{D.}~\bibnamefont{Huffman}},
  \emph{\bibinfo{title}{Absorption and scattering of light by small particles}}
  (\bibinfo{publisher}{John Wiley and Sons, Inc., New York},
  \bibinfo{year}{1983}).

\bibitem[{\citenamefont{Moroz}(2009)}]{JOSAB_26_517_moroz}
\bibinfo{author}{\bibfnamefont{A.}~\bibnamefont{Moroz}}, \bibinfo{journal}{J.
  Opt. Soc. Am. B} \textbf{\bibinfo{volume}{26}}, \bibinfo{pages}{517}
  (\bibinfo{year}{2009}).

\bibitem[{\citenamefont{Jensen et~al.}(1999)\citenamefont{Jensen, Kelly,
  Lazarides, and Schatz}}]{JCS_10_295_schatz}
\bibinfo{author}{\bibfnamefont{T.}~\bibnamefont{Jensen}},
  \bibinfo{author}{\bibfnamefont{L.}~\bibnamefont{Kelly}},
  \bibinfo{author}{\bibfnamefont{A.}~\bibnamefont{Lazarides}},
  \bibnamefont{and} \bibinfo{author}{\bibfnamefont{G.~C.}
  \bibnamefont{Schatz}}, \bibinfo{journal}{J. Cluster Sci.}
  \textbf{\bibinfo{volume}{10}}, \bibinfo{pages}{295} (\bibinfo{year}{1999}).

\bibitem[{\citenamefont{Zou and Schatz}(2004)}]{JCP_121_12606_schatz}
\bibinfo{author}{\bibfnamefont{S.}~\bibnamefont{Zou}} \bibnamefont{and}
  \bibinfo{author}{\bibfnamefont{G.~C.} \bibnamefont{Schatz}},
  \bibinfo{journal}{J. Chem. Phys.} \textbf{\bibinfo{volume}{121}},
  \bibinfo{pages}{12606} (\bibinfo{year}{2004}).

\bibitem[{\citenamefont{Markel}(2005{\natexlab{a}})}]{JPhysB_38_L115_markel}
\bibinfo{author}{\bibfnamefont{V.~A.} \bibnamefont{Markel}},
  \bibinfo{journal}{J. Phys. B} \textbf{\bibinfo{volume}{38}},
  \bibinfo{pages}{L115} (\bibinfo{year}{2005}{\natexlab{a}}).

\bibitem[{\citenamefont{Augui\'e and Barnes}(2009)}]{OL_34_401_barnes}
\bibinfo{author}{\bibfnamefont{B.}~\bibnamefont{Augui\'e}} \bibnamefont{and}
  \bibinfo{author}{\bibfnamefont{W.~L.} \bibnamefont{Barnes}},
  \bibinfo{journal}{Opt. Lett.} \textbf{\bibinfo{volume}{34}},
  \bibinfo{pages}{401} (\bibinfo{year}{2009}).

\bibitem[{\citenamefont{Persson and Liebsch}(1983)}]{PRB_28_4247_persson}
\bibinfo{author}{\bibfnamefont{B.~N.~J.} \bibnamefont{Persson}}
  \bibnamefont{and} \bibinfo{author}{\bibfnamefont{A.}~\bibnamefont{Liebsch}},
  \bibinfo{journal}{Phys. Rev. B} \textbf{\bibinfo{volume}{28}},
  \bibinfo{pages}{4247} (\bibinfo{year}{1983}).

\bibitem[{\citenamefont{Hinrichsen et~al.}(1986)\citenamefont{Hinrichsen,
  Feder, and J\o\/ssang}}]{JStatPhys_44_793_feder}
\bibinfo{author}{\bibfnamefont{E.~L.} \bibnamefont{Hinrichsen}},
  \bibinfo{author}{\bibfnamefont{J.}~\bibnamefont{Feder}}, \bibnamefont{and}
  \bibinfo{author}{\bibfnamefont{T.}~\bibnamefont{J\o\/ssang}},
  \bibinfo{journal}{J. Stat. Phys.} \textbf{\bibinfo{volume}{44}},
  \bibinfo{pages}{793} (\bibinfo{year}{1986}).

\bibitem[{ZZZ()}]{ZZZ_cc_fitting}
\bibinfo{note}{The fitted parameters are $a_{0}=0.5055$, $a_{1}=1.445$,
  $a_{2}=-1.445$, $b_{0}=3.619$, $b_{1}=0.003095$, $b_{2}=16.64$,
  $b_{3}=-1.016$, $c=0.991$, $d_{0}=2.019$, $d_{1}=1.185$.}

\bibitem[{\citenamefont{Zou et~al.}(2004)\citenamefont{Zou, Janel, and
  Schatz}}]{JCP_120_10871_schatz}
\bibinfo{author}{\bibfnamefont{S.}~\bibnamefont{Zou}},
  \bibinfo{author}{\bibfnamefont{N.}~\bibnamefont{Janel}}, \bibnamefont{and}
  \bibinfo{author}{\bibfnamefont{G.~C.} \bibnamefont{Schatz}},
  \bibinfo{journal}{J. Chem. Phys.} \textbf{\bibinfo{volume}{120}},
  \bibinfo{pages}{10871} (\bibinfo{year}{2004}).

\bibitem[{\citenamefont{Markel}(2005{\natexlab{b}})}]{JCP_122_097101_markel}
\bibinfo{author}{\bibfnamefont{V.~A.} \bibnamefont{Markel}},
  \bibinfo{journal}{J. Chem. Phys.} \textbf{\bibinfo{volume}{122}},
  \bibinfo{pages}{097101} (\bibinfo{year}{2005}{\natexlab{b}}).

\bibitem[{\citenamefont{Zou and Schatz}(2005)}]{JCP_122_097102_schatz}
\bibinfo{author}{\bibfnamefont{S.}~\bibnamefont{Zou}} \bibnamefont{and}
  \bibinfo{author}{\bibfnamefont{G.~C.} \bibnamefont{Schatz}},
  \bibinfo{journal}{J. Chem. Phys.} \textbf{\bibinfo{volume}{122}},
  \bibinfo{pages}{097102} (\bibinfo{year}{2005}).

\bibitem[{\citenamefont{Langhammer et~al.}(2007)\citenamefont{Langhammer,
  Kasemo, and Zori\'c}}]{JCP_126_194702_langhammer}
\bibinfo{author}{\bibfnamefont{C.}~\bibnamefont{Langhammer}},
  \bibinfo{author}{\bibfnamefont{B.}~\bibnamefont{Kasemo}}, \bibnamefont{and}
  \bibinfo{author}{\bibfnamefont{I.}~\bibnamefont{Zori\'c}},
  \bibinfo{journal}{J. Chem. Phys.} \textbf{\bibinfo{volume}{126}},
  \bibinfo{pages}{194702} (\bibinfo{year}{2007}).

\end{thebibliography}

\end{document}